\documentclass[]{aa} 

\newcommand{\msun}{$\mathrm{M_{\odot}}$}
\newcommand{\teff}{$T_{\rm eff}$}

\usepackage{natbib,twoopt}
\usepackage[breaklinks=true,colorlinks=true,linkcolor=black,citecolor=blue]{hyperref}
\bibpunct{(}{)}{;}{a}{}{,}
\usepackage[varg]{txfonts}
\usepackage{graphicx}
\usepackage[utf8]{inputenc}
\usepackage{color}
\usepackage{epsfig}
\usepackage{longtable}
\usepackage{booktabs}
\usepackage{tabularx}

\begin{document}

\title{Mass effect on the lithium abundance evolution of open clusters: Hyades, NGC 752, and M67}
 \subtitle{}

   \author{M. Castro\inst{1}     
           \and T. Duarte\inst{1} 
           \and G. Pace\inst{2}
           \and J. -D. do Nascimento Jr.\inst{1,3}
}

   \offprints{M. Castro, email: mcastro@dfte.ufrn.br}

\institute{Departamento de Física Teórica e Experimental, Universidade Federal do Rio Grande do Norte, CEP: 59072-970 Natal, RN, Brazil
\and Instituto de Astrofísica e Ciência do Espaço, Universidade do Porto, Rua das Estrelas, 4150-762 Porto, Portugal
\and Harvard-Smithsonian Center for Astrophysics, Cambridge, MA 02138, USA
}

\date{Received : Accepted : }

\abstract{Lithium abundances in open clusters provide an effective way of probing mixing processes in the interior of solar-type stars and convection is not the only mixing mechanism at work. To understand which mixing mechanisms are occurring in low-mass stars, we test non-standard models, which were calibrated using the Sun, with observations of three open clusters of different ages, the Hyades, NGC 752, and M67. We collected all available data, and for the open cluster NGC 752, we redetermine the equivalent widths and the lithium abundances. Two sets of evolutionary models were computed, one grid of only standard models  with microscopic diffusion and one grid with rotation-induced mixing, at metallicity [Fe/H] = 0.13, 0.0, and 0.01 dex, respectively, using the Toulouse-Geneva evolution code. We compare observations with models in a color-magnitude diagram for each cluster to infer a cluster age and a stellar mass for each cluster member. Then, for each cluster we analyze the lithium abundance of each star 
as a function of mass. 
The data for the open clusters Hyades, NGC 752, and M67, are compatible with lithium abundance being a function of both age and mass for stars in these clusters. Our models with meridional circulation qualitatively reproduce  the general trend of lithium abundance evolution as a function of stellar mass in all three clusters. This study points out the importance of mass dependence in the evolution of lithium abundance as a function of age. Comparison between models with and without rotation-induced mixing shows that the inclusion of meridional circulation is essential to account for lithium depletion in low-mass stars. However, our results suggest that other mechanisms should be included to explain the Li-dip and the lithium dispersion in low-mass stars.}


\titlerunning{Mass effect on the lithium abundance evolution of open clusters}
\authorrunning{Castro et al.}

\maketitle

\section{Introduction}
\label{sec:Intro}

Lithium ($^7$Li) is a fragile element that is destroyed at temperatures above $\sim$2.5 $\times~10^6$ K. The lithium depletion in stars depends on several factors such as mass, age, metallicity, rotation, magnetic fields, mass loss, and mixing mechanisms \citep[e.g.][]{dm84,dp97,ventura98,charbonneltalon05}. Abundance measurements of this element in open cluster stars allow us to investigate mixing in stellar interior and to put constraints on the non-convective mixing processes as a function of mass, age, rotation, and metallicity for coeval stars. To understand the mixing mechanisms and to explain the complex evolution of stellar chemical elements, we need to constrain the effect of mass. As a part of this complexity, several unexpected results related to the depletion of lithium in stars have been found and reveal our limited understanding of the physics acting in the stars' interiors. One of these unexpected results concerns the measurements of lithium abundance in late-F and early G-type stars, 
showing empirical evidence that is in contradiction with the standard model predictions, which only include convection. For stars in these spectral classes, authors have shown lithium depletion during the main sequence \citep{BT86,boesgaard87,ph88}. However, convective zones of low-mass main sequence (MS) stars do not reach regions in the stellar interior, where the temperature is sufficiently high to drive lithium destruction \citep[see e.g.][ and references therein]{randich06}. This depletion is mass-age dependent \citep{donascimento09,melendez10,pace12}. For  stars of one solar mass, including the Sun and those stars, which are spectroscopically indistinguishable from the Sun called solar twins \citep{cayrel96}, the extremely low lithium abundance of about one hundred times lower than its original value as measured in meteorites, represents a long-standing puzzle.

During the pre-main-sequence (PMS), studies showed that Li destruction in stars with mass that is equal or larger to the solar one should be weak \citep{martin93,jones97}. \citet{randich97} determined the lithium abundance in 28 stars of the 28 Myr old cluster IC 2602. They compared the lithium abundance distribution as a function of stellar mass of this sample with those of the $\alpha$ Persei (50 Myr old) and the Pleiades (70 Myr old) clusters, and showed that only latest-type stars present a significant lithium depletion during the PMS. Early-K and G-type stars have a slightly lower lithium abundance in the oldest clusters, and F-type stars do not present any significant lithium depletion.

According to the standard stellar models, lithium depletion should be a unique function of mass, age, and chemical composition. Stars with close effective temperature in a given cluster should have undergone the same amount of lithium depletion. In this context, lithium abundances have been studied extensively in open clusters of different ages and revealed a complex behavior with an important scattering \citep[see e.g.][ and references therein]{pinsonneault97}. The star-to-star scatter in lithium abundance exists among solar-type stars of the $\sim$2 Gyr old cluster NGC 752, as well as in the solar age, solar metallicity open cluster M67 \citep[e.g.][]{pasquini97,jones99}. From the study of the open cluster NGC 752  by \citet{hobbs86a}, \citet{ph88}, and  \citet{balachandran95}, and from the 4 Gyr old open cluster M67 studied by \citet{hobbs86b}, \citet{spite87}, and \citet{garcialopez88}, authors conclude about a significant depletion relative to younger systems that increases with age. 
Despite decades of observations in clusters  and field stars, the mechanism that drives the evolution of the surface lithium abundance and its scattering in main-sequence field stars and stars belonging to open clusters is far from being fully understood, and the lithium depletion expected from theoretical models needs to be adjusted to reproduce the mass and age-dependence. As for the open cluster M67 \citep{pace12}, we explore  this mass-age dependence and the star-to-star scatter at the same mass and different ages, with particular attention to the quality of the spectroscopy behind the lithium  abundance determination. In this work, we continue to homogenize previous studies on open clusters by obtaining a consistent set of temperatures and masses, and by applying corrections to the lithium abundances when needed, due to revised effective temperatures, by taking into account non-LTE effects on the lithium abundances. We analyze stars from the main-sequence to the red giant branch, comparing our models to the observed behavior of lithium abundance as a function of mass and age.

This paper is organized as follows. In Sect. \ref{sec:Obs}, we describe the working sample of the three clusters. In  Sect. \ref{sec:Models},  we present  details of our evolutionary stellar models, and  in  Sect. \ref{sec:cmd} we construct the color-magnitude diagrams of the open clusters.  In  Sect. \ref{sec:predictions}, we discuss  the evolution of the lithium abundance predicted from our models. In Sect. \ref{sec:comparisons}, we compare the models with the observations from the studied clusters. Finally, in  Sects. \ref{sec:discussion} and  \ref{sec:conclusions}, we present our discussion and conclusions.

\section{Working sample}
\label{sec:Obs}

Our working sample is composed of lithium non-LTE abundance measurements for 67 F-G dwarfs in the Hyades from \citet{takeda13}. \citeauthor{takeda13} re-studied 49 stars with lithium abundance measurements that are available in the literature \citep{randich07,thorburn93,soderblom90,boe88,BT86,duncan83,rebolo88}, and added 19 more stars to their sample. This sample is currently the largest  for lithium abundances of Hyades stars. We decided not to use the star HD 285252 of the original sample because its lithium abundance is very low and uncertain ($A({\rm Li})  = -0.29$).
\citet{takeda13} do not directly provide  uncertainty in their lithium abundance measurements, but they do provide all necessary data in electronic format
to compute it. The uncertainty on lithium abundance  reported in Table \ref{tab:interphyades} was obtained by adding its three
components quadraticall , i.e. that originated from the error of temperature, gravity, and microturbolence,
respectively. Each term was defined as $1/2(A(\mathrm{Li},P-\Delta P) - A(\mathrm{Li},P+\Delta P))$, where $P$ is the parameter considered (temperature, gravity, or microturbulence), $\Delta P$ its uncertainty according to \citet{takeda13}, which assumes $\Delta T = 100 \ K$, and the two other  parameters  were kept constant. The temperature is the only significant source of error for the lithium abundance.

Masses and stellar parameters of the 67 sample stars were computed for each star by taking the closest isochrone point to the stellar data-point on the color-magnitude diagram (see Sect. \ref{sec:cmd}). The isochrone was chosen by imposing [Fe/H] = 0.13 \citep{paulson03} and selecting the best fitting  age, which was found to be 0.79 Gyr. B-V colours and absolute V magnitude are taken from \citet{takeda13}, who in turn used photometry and parallaxes for individual Hyades members from the Hipparcos catalogue \citep{debruijne, esa97}. Lithium abundances for the individual Hyades stars are directly derived by \citet{takeda13}, and  do not take into account this adjustment on the stellar parameters, which was applied by projection on the isochrone.

Takeda's lithium abundance measurements are already accurate and homogeneous. Revisiting them, as we did for NGC752 stars,  would not change the conclusions of this paper, so we did not. The calculation of stellar parameters for Hyades stars was mainly intended to derive stellar masses, which we chose as an independent variable to examine the trend of lithium abundances, as we did for the other clusters.\\

Equivalent width (EW) measurements of the lithium line at 6708 {\AA} for NGC 752 members were taken from the following sources: \citet{sestito04,hobbs86a,ph88}. \citet{psh88} give lithium abundances (mostly upper limits) for 11 giants, which were transformed into (upper limits of) EWs by reverting the LTE analysis using the \citet{lind09} code. Since \citeauthor{psh88} do not mention what metallicty they adopted, we assumed a solar  metallicity for the stars. There are three stars in common between the older works and that of \citeauthor{sestito04} In these cases, the value from the latter work is adopted. The mean difference between the EWs that is published in the two works for these three stars in common is used as an estimation of the errors on the EW for the other measurements given in \citet{hobbs86a}, because they are not provided by the authors. The stars in common are PLA701, PLA859, and PLA921 (H146, H207, and H229, respectively, in the numbering system by \citet{heinemann26}), and the couple of measurements are, respectively, 42 and 45, 26.2 and 7, and 54 and 63. The  values are in m$\AA$, the result from \citeauthor{sestito04} being the first of each pair. The average difference, after rounding, is 10 m$\AA$, which we adopted as error on EW for all stars for which only the old measurements are available. The EWs collected were used to compute abundances with the code by K. Lind, employing the full non-LTE analysis and the stellar parameters computed, as described below.

Errors on lithium abundances were calculated by adding quadratically its two main components, i.e. that due to the equivalent width and that due to the temperature. For the former, we simply computed: $dA(\mathrm{Li})_{EW}=1/2(A(\mathrm{Li}, EW+\Delta_{EW})-A(\mathrm{Li}, EW-\Delta_{EW}))$, where temperature, gravity, and metallicity were kept constant. To estimate the error due to the uncertainty in the temperature, we computed each stellar temperature in two different  ways. The most accurate value was drawn from the isochrone in the CMD, as described in more detail below. We will refer to it as $T_{\mathrm{iso}}$. The other value,
$T_{\mathrm{phot}}$, was obtained from the photometric calibration of \citet{casagrande10}, or \citet{kucinskas05} for the giants, and the photometry in the  reference papers. We then compute, for each dwarf, the following: $dA(\mathrm{Li})_{T}=A(\mathrm{Li}, T_{\mathrm{iso}})-A(\mathrm{Li}, T_{\mathrm{phot}})$.

The mean cluster iron content estimates range from $[\mathrm{Fe/H}] = -0.09$ dex \citep[][based on 8 dwarfs]{ht92} to $[\mathrm{Fe/H}] = 0.08 \pm 0.04$ \citep[][based on 4 giants]{carrera&pancino11}. The other two studies (\citealt{sestito04}, based  on 18 dwarfs, and \citealt{reddy}, based on 4 giants) favour a substantially solar metallicity ($[\mathrm{Fe/H}] = 0.01 \pm 0.04$ and $[\mathrm{Fe/H}] = -0.02 \pm 0.05$, respectively). In conclusion, the range of possible values for the cluster metallicity is aproximately $[\mathrm{Fe/H}] = 0.0 \pm 0.1$.

As we did for the Hyades, we computed stellar parameters for each star (effective temperature and mass) from the isochrone point closest to the stellar data point on the color-magnitude diagram. We adopted the best fitting isochrone, with an age of 1.59 Gyr for $[\mathrm{Fe/H}] = 0.0$ (see Sect. \ref{sec:cmd}). \\

\begin{figure}[t!]
 \begin{center}
 \vspace{-0.1in}
  \includegraphics[width=9 cm,height=9cm]{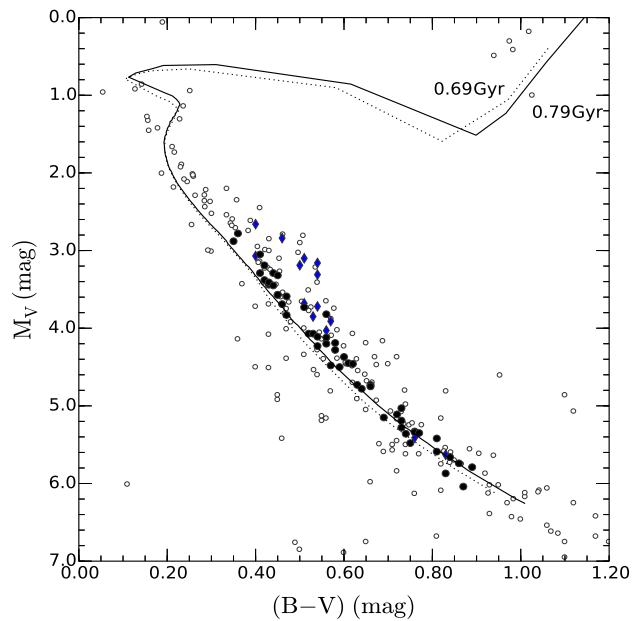}
 \end{center}
 \vspace{-0.2in}
 \caption{Color-magnitude diagram of the Hyades open cluster. Open circles are the whole sample of Hyades stars with magnitudes from the Hipparcos catalogue. Filled black circles are the sample of stars form \citet{takeda13} for which we have a lithium abundance estimation. Blue diamonds represent the known binaries of this sample. The continuous line is the isochrone of 0.79 Gyr, calculated from our TGEC models with extra mixing, and the dotted line is the isochrone of 0.69 Gyr, calculated from standard models.}
 \label{fig:cmd_hyades}
\end{figure}

\begin{figure}[t!]
 \begin{center}
 \vspace{-0.1in}
  \includegraphics[width=9 cm,height=9cm]{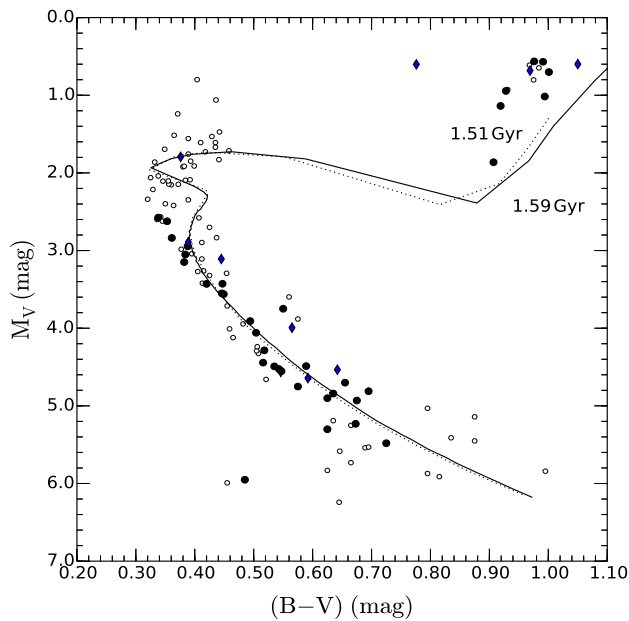}
 \end{center}
 \vspace{-0.2in}
 \caption{Color-magnitude diagram of the open cluster NGC 752. Open circles are the whole sample of NGC 752 stars from \citet{daniel94}. Filled black circles are our working sample of stars for which we have a lithium abundance estimation. Blue diamonds represent the known binaries of this sample. The continuous line is the isochrone of 1.59 Gyr calculated from our TGEC models with extra mixing and the dotted line is the isochrone of 1.51 Gyr calculated from standard models.}
 \label{fig:cmd_ngc752}
\end{figure}

The stars of both samples of the Hyades and NGC 752 clusters were all supposed to be single stars when included in the reference studies. When we checked the data using the study of \citet{vanleeuwen07}, queried through WEBDA, it turned out that these samples were contaminated by binaries. The contamination is about 20\% for the Hyades cluster (14 out of 67) and for NGC 752 (10 out of 46). All known binaries in both samples are identified in Tables \ref{tab:interphyades} and \ref{tab:interpngc752} and in Figs. \ref{fig:cmd_hyades}, \ref{fig:cmd_ngc752}, \ref{fig:li-mass_hyades}, and \ref{fig:li-mass_ngc752}.\\

The data relative to lithium abundance for the M67 cluster, corrected for non-LTE effects using the grid of \citet{lind09}, were analyzed in \citet{pace12}, to whom we redirect the reader.

\section{Stellar evolutionary models}
\label{sec:Models}

We present here the models computed using the Toulouse-Geneva stellar evolution code (TGEC) \citep{huibonhoa08,donascimento09}. We used the same input physics as in \citet{pace12}. In particular, the transport of chemical elements and angular momentum is driven by different processes, which we briefly describe hereafter.

In the outer layers, convection is modeled according to the \citet{bohm58} formalism of the mixing-length theory. The mixing-length parameter $\alpha = l/H_p$, where $l$ is the characteristic mixing length and $H_p$ is the pressure scale height, is a free parameter in our models. Below the convective zone, we introduce a convective undershooting with a depth of 0.09 $H_p$ so that, for the Sun, the combined mixing reaches the depth deduced by helioseismology \citep[$r_{cz}/R_{\odot}=0.713 \pm 0.001$, ][]{basu&antia97}.
 
In the radiative zone, microscopic diffusion, which is the process of element segregation by gravitational and thermal diffusion \citep{eddington16,chapman17}, is treated by computing convergent series of a maxwellian distribution function, which is the solution at equilibrium of the Bolzmann equation for dilute collision-dominated plasmas, with sucessive approximations \citep[see the Chapman-Enskog procedure described in][]{chapman&cowling70}. For collisons between charged ions, we use the \citet{paquette86} method, which introduces a screened coulomb potential. Microscopic diffusion is essential in stellar models to account for abundances anomalies in Ap and Am stars \citep{michaud70} and to improve the consistency with helioseismology \citep{cox89,bahcall95,cd96,richard96,turcotte98,schlattl02}.
 
When a star rotates, centrifugal effects modify the gravity equipotencials, which are no longer spherical. It induces a macroscopic circulation of matter in the radiative zone between polar and equadorial regions, the so-called meridional circulation \citep{eddington26,sweet50}. \citet{zahn92} suggested that the meridional flow induces a transport of angular momentum, creating shears which become unstable in the horizontal direction, while the vertical shears are stabilized by the density gradient. The coupling between the meridional advection and the horizontal turbulence leads to an anisotropic mixing of the chemical species, parameterized as an effective diffusion coefficient. \citet{mestel&moss86} showed that the nuclear-induced $\mu$-gradients slowly stabilize the circulation and expel it from the core toward the external layers. \citet{vt03} show that this feedback effect, which is due to the $\mu$-gradients, strongly modifies the meridional circulation and meanwhile reduces the efficiency of diffusion.
 Following \citet{zahn92}, \citet{talon&zahn97}, and \citet{maeder&zahn98}, the rotational mixing is computed by a diffusion coefficient that we write as

\begin{center}
 \begin{equation}
  D_{turb} = \alpha_{turb} r | U_{\rm r} |
 \end{equation}
\end{center}

with 

\begin{center}
 \begin{equation}
  \alpha_{\rm turb} = C_{\rm v} + \frac{1}{30 C_{\rm h}}
 \end{equation}
\end{center}  
  
where $C_{\rm v}$ and $C_{\rm h}$ are unknown parameters relating to the vertical and horizontal part of the shear-induced anisotropic turbulent coefficient that is due  to the transport of angular momentum, respectively. According to the assumption of strong anisotropic turbulence, they must satisfy the condition: $C_{\rm v} << C_{\rm h}$. $\mathbf{U_{\rm r}}$ is the vertical velocity amplitude which includes a classical meridional circulation term, which is directly linked to the rotation velocity $\Omega$, a $\mu$-gradient-induced term, a term related to the time variations of the differential rotation, and a term that appears in the case of large horizontal turbulence \citep[see][ for more details]{tv03b}.

At the bottom of the convective zone, the transition layer between the differential rotation of the convective zone and the radiative interior rotating as a solid body is called the tachocline. This shear layer undergoes a strong anisotropic turbulence, with much stronger viscous transport in the horizontal than in the vertical direction, reducing the differential rotation and inhibiting its spread deep inside the radiative interior \citep{spiegel&zahn92}. We assume that the absolute size of the tachocline is constant with time. \citet{brun98} showed it can be modeled by an exponential diffusion coefficient added to the turbulent diffusion coefficient. Thus, we use an effective diffusion coefficient:

\begin{center}
 \begin{equation}
  D_{eff} = D_{turb} + D_{tacho}
 \end{equation}
\end{center}

with

\begin{center}
 \begin{equation}
  D_{tacho} = D_{bcz} \exp \left( \ln 2 . \frac{r - r_{bcz}}{\Delta} \right)
 \end{equation}
\end{center}

where $D_{\rm bcz}$ and $r_{\rm bcz}$ are, respectively, the diffusion coefficient and the radius at the base of the convective zone, and $\Delta$ is the half width of the tachocline \citep{richard04}. This is the effective diffusion coefficient that appears in the equation of the chemical transport for the mean concentration of the different species.\\

We calibrated our models with the Sun as in \citet{richard04}. There is no a priori reason why the calibration of the parameters for the model of 1.00 \msun \ should also hold  for different masses and evolutionary times but, given that no other standard calibrators are available, our approach is the best possible one for now.
The calibration procedure of our models is the same as described in Sect. 4.2 in \citet{pace12}. The mixing-length parameter $\alpha$ and the initial helium abundance $Y_{\rm ini}$ enable us to calibrate the present evolutionary status of the Sun. By triangulation, we found the best couple $(\alpha,Y_{\rm ini})$ that allows us to reproduce the solar radius and the solar luminosity at the solar age \citep[$L_{\odot} = 3.8515 \pm 0.0055 \times 10^{33}$ erg.s$^{-1}$ and $R_{\odot} = 6.95749 \pm 0.00241 \times 10^{10}$ cm at $t_{\odot} = 4.57 \pm 0.02$ Gyr][]{richard04}. For the best-fit solar model, we obtained $L = 3.8501 \times 10^{33}$ erg.s$^{-1}$ and $R = 6.95524 \times 10^{10}$ cm at an age $t = 4.57$ Gyr with $\alpha = 1.69$ and $Y_{\rm ini} = 0.268$. The rotation-induced mixing efficiency (meridional circulation and turbulent motions) is calibrated using the parameters $C_{\rm h}$ and $\alpha_{\rm turb}$. $C_{\rm v}$ is determined from these two parameters through Eq. 2. As explained in \citet{richard04}, the feedback currents, which are due to $\mu$-gradients, are strongly dependent on $C_{\rm h}
$, which is directly related to the horizontal turbulence. A strong horizontal turbulence tends to homogenize the horizontal layers and thereby smooths the horizontal $\mu$-gradients. In this case, the small induced $\mu$-currents may not be able to compensate for the currents because of classical meridional circulation and the mixing may remain efficient during a long timescale. In particular, we intend to put a mixing that is efficient and deep enough to smooth the diffusion-induced helium gradient, which lies below the surface convective zone, thus improving the agreement between the model and seismic sound-speed profiles. On the other hand, weak horizontal turbulence leads to important $\mu$-currents, which strongly reduce the mixing. In our models, this had to be weak and shallow enough to avoid the destruction of beryllium. We used the values $C_{\rm h}$ = 9000 and $\alpha_{\rm turb}$ = 1, and obtained an excellent agreement with helioseismology, which was more accurate than 1\% for most of the stars in terms of sound velocity, except in the 
deep interior, where the discrepancy reaches 1.5\%. We also obtained a slight destruction of beryllium by a factor of 1.17 with respect to the meteoritic value, which is well within the error in the determination of the solar beryllium abundance following \citet{gs98}. Beryllium destruction in low-mass stars will be subject of a further study. Finally, the tachocline is calibrated using the parameters $D_{\rm bcz}$ and $\Delta$ to obtain the solar lithium abundance at the solar age. Helioseismic constraints indicate that the solar tachocline thickness, defined as the region where the effective diffusion coefficient
$D_{tacho}$ increases from 0.08 to 0.92 of its maximum value $D_{bcz}$, should be lower than 0.04 $R_{\rm \odot}$ \citep{charbonneau99}, leading to a $\Delta$ lower than 0.01135 $R_{\rm \odot}$ \citep{richard04}. Using $D_{\rm bzc} = 2.02 \times 10^{5}$ m$^{2}$.s$^{-1}$ and $\Delta = 0.60 \times 10^{9}$ cm = 0.0086 $R_{\rm \odot}$, we obtained a surface lithium content relative to the initial one ${\rm Li}/{\rm Li}_0 = 0.60 \times 10^{-2}$. To obtain the lithium abundance $A({\rm Li})$, the question of the initial abundance on the 
ZAMS is important. Stars with different masses in the range 0.8 - 2.0 \msun do not suffer the same lithium depletion during the PMS. In the Sun, this depletion should have been low and, therefore, we kept following the calibration procedure from \citet{richard96,richard04}, which neglects
the lithium destruction during the PMS in solar models. Consequently, we adopted the meteoritic lithium abundance from \citet{asplund09} $A({\rm Li})_0 = 3.26$ for the initial abundance, and obtained at the solar age $A({\rm Li}) = 1.04$, which agrees well with solar lithium abundance measurements from \citet{asplund09} ($A({\rm Li}) =  1.05 \pm 0.10$).

The three steps of the calibration are not completely independent as the variations in the rotation-induced mixing parameters slightly modify the luminosity and radius of the model. Some iterations are necessary to obtain the best-fit solar model.\\

We computed two grids of evolutionary models for each cluster, one grid of rotation-induced mixing, and another grid of standard models (with microscopic diffusion), to check the influence of the rotation-induced mixing on the lithium abundance depletion. For the Hyades, we computed models with a range of masses from 0.80 to 2.25 \msun\ with a step of 0.01 \msun\, and with a metallicity of ${\rm [Fe/H]} = 0.13$. For NGC 752, we computed models with masses from 0.80 to 1.77 \msun\ with a step of 0.01 \msun\, with a metallicity of ${\rm [Fe/H]} = 0.0$. Finally, for M67, our models have been computed with masses in the range 0.90 to 1.34 \msun, with a step of 0.01 \msun, and with a metallicity of ${\rm [Fe/H]} = 0.01$, which is a simple average of different estimates for the metallicity of M67 \citep[see][]{pasquini08}. We ran the models from the zero age main sequence (ZAMS) to the top of the red giant branch (RGB) for the most massive stars. The input parameters for all the models 
are the same as for the 1.00 \msun\ model. In particular, we used the same initial lithium abundance for all masses. For solar mass and larger, we can neglect lithium depletion during the PMS, and the choice of the meteoritic lithium abundance since initial abundance of the models is straightforward. For lower masses than the solar one, initial lithium abundance should be lower since the lithium depletion during the PMS increases with decreasing mass. However, it is a tricky task to choose a value for the lithium abundance at the ZAMS as a function of stellar mass. \citet{dantona94} determined that, depending on the physics input of convection treatment and opacities, a model with 0.8 \msun \ presents a lithium depleted of between 1.7 and 3.3 dex, at the age of $\alpha$ Persei. The smallest destruction ($\sim 1.7$ dex) is the most consistent with lithium abundance observations of $\alpha$ Per and corresponds to models with the \citet{kurucz91} opacities and \citet{canuto&mazzitelli90} convection treatment. In this case, we chose to adopt the meteoritic lithium abundance as initial abundance as well, keeping in mind that, for these masses, it should be lower.

\section{Color-magnitude diagrams}
\label{sec:cmd}

Accurate determinations of cluster ages and the masses of their members are an essential requirement when comparing our models to the cluster observations and to study the evolution with stellar mass of lithium abundance in each cluster. The classic way to obtain both is to compare model isochrones with data points on the color-magnitude diagram, as in \citet{pace12}.

To construct the color-magnitude diagram of the open cluster Hyades presented in Figure \ref{fig:cmd_hyades}, we used a reddening $E(B - V) = 0.001$ mag \citep{taylor06}, a distance modulus $(m - M)_0 = 3.35$ mag \citep{mcArthur11,perryman98}, and a metallicity $[\mathrm{Fe/H}] = +0.13$ \citep{paulson03}. 
We plot the whole sample of Hyades stars with magnitudes from the Hipparcos catalogue along with our working sample of stars form \citet{takeda13} for which we have lithium abundance estimation. Blue diamonds represent the known binaries of this sample.
For the isochrones, we computed two grids of models, as presented in Sect. \ref{sec:Models}. Matching the position of the turn-off hook of the isochrones with the left envelope of the observations, we estimated an age of 0.69 Gyr with the grid of standard models, and 0.79 Gyr with the grid of models with extra mixing, which falls right in between the classical estimation of 625 Myr \citep{perryman98} and that by \citet{brandtHuang15} of 950 Myr. The data points that are found significantly far away from the isochrone could be due to photometric errors, non-members, or binaries. We can see that the known binaries are mostly off the isochrone. Also, a non-detected companion in a binary system would make the star appear  more luminous and colder than it is, therefore a bias could exist in the CMD that moves some points up and/or to the left. However, in the case of the Hyades, this effect should be very small as the cluster has been very well studied.

The color-magnitude diagram for NGC 752 is presented in Figure \ref{fig:cmd_ngc752}. We built it using the grids of models exposed in Sect. \ref{sec:Models}. Then, we shifted the resulting isochrone according to a reddening of $E(B - V) = 0.035$ mag and using a distance modulus of $(m - M)_0 = 8.25$ mag \citep{daniel94}. From the position of the turn-off hook of the isochrones, we estimated an age of 1.51 Gyr with the grid of standard models, and 1.59 Gyr with the grid of models with extra mixing. The positions of the turn-off hook and the giant branch of the isochrone depend on the input physics in the models such as metallicity, diffusion, and overshooting at the convective core. In particular, the cut-off of the diffusion when the model enters into the RGB or a bad calibration of the overshooting, could explain why the giants are off the isochrone.

\begin{figure}[t!]
 \begin{center}
 \vspace{-0.1in}
  \includegraphics[width=9 cm,height=9cm]{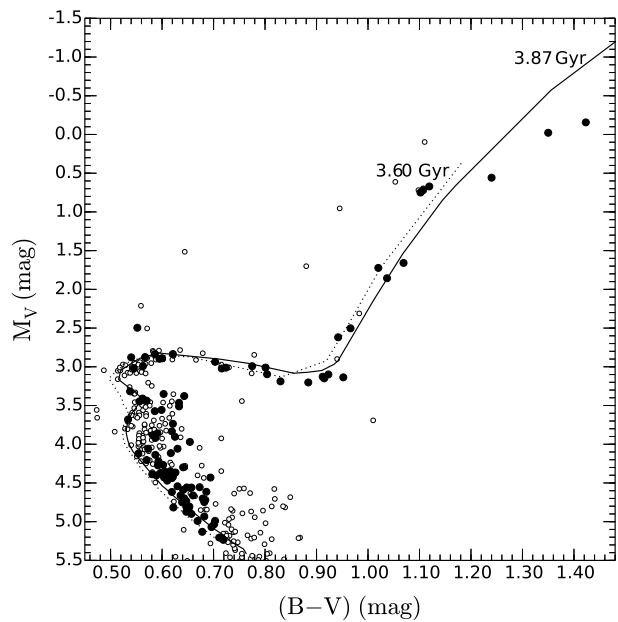}
 \end{center}
 \vspace{-0.2in}
 \caption{Color-magnitude diagram of the open cluster M67. The black filled circles represent our work sample, i.e., stars with lithium abundance measurements. Stars with photometry available from \cite{yadav08}, but with no lithium abundances, are displayed as open circles. The continuous isochrone from TGEC models with extra mixing corresponds to an age of 3.87 Gyr. The dotted line is the isochrone of 3.60 Gyr calculated from standard models.}
 \label{fig:cmd_m67}
\end{figure}

For the open cluster, M67, see the paper \citet{pace12}, where model calibration and mass determination, a color-magnitude diagram, and an estimation of the cluster age of 3.87 Gyr have already been presented. In the present work, we added an isochrone that is made up of standard models to the color-magnitude diagram of M67 and found an age of 3.60 Gyr. The new plot, including this isochrone, is presented in Fig. \ref{fig:cmd_m67}.

\section{Predicted lithium evolution as a function of mass and age}
\label{sec:predictions}

\begin{figure}[t!]
 \begin{center}
 \vspace{-0.1in}
  \includegraphics[width=9 cm,height=9 cm]{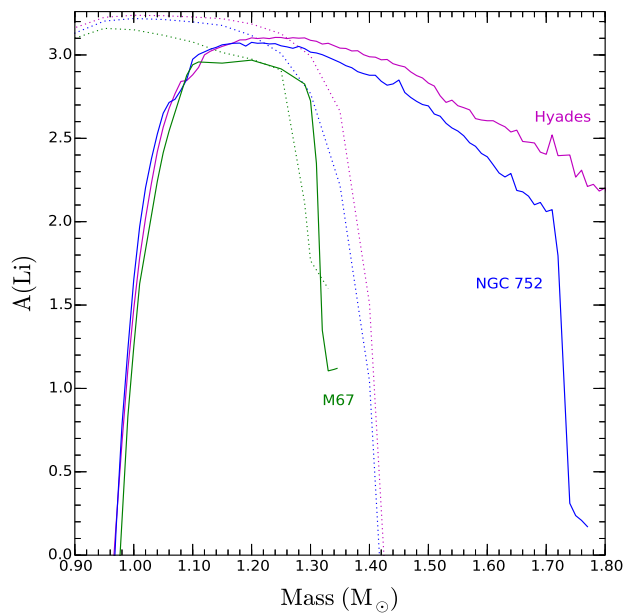}
 \end{center}
 \vspace{-0.2in}
 \caption{Lithium abundance as a function of mass for isochrones with ages estimated for each cluster. Contiunuous isochrones are from models with extra mixing and dotted isohrones from standard models. Magenta lines represent the isochrones for the Hyades with $[\mathrm{Fe/H}] = +0.13$ and ages of 0.79 Gyr and 0.69 Gyr, respectively; blue lines represent the isochrones for NGC 752 with $[\mathrm{Fe/H}] = +0.00$ and ages of 1.59 Gyr and 1.51 Gyr, respectively; and green lines represent the isochrones for M67 with $[\mathrm{Fe/H}] = +0.01$ and ages of 3.87 Gyr and 3.60 Gyr, respectively.}
 \label{fig:li-mass_models}
\end{figure}

\begin{figure}[t!]
 \begin{center}
 \vspace{-0.1in}
  \includegraphics[width=9 cm,height=9 cm]{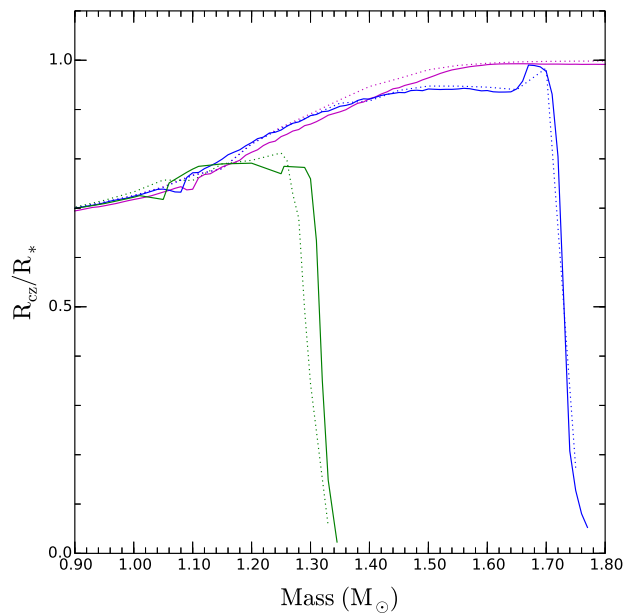}
 \end{center}
 \vspace{-0.2in}
 \caption{Relative inner convective zone radius as a function of stellar mass for isochrones with ages estimated for each cluster. The line types and colors are the same as in Fig. \ref{fig:li-mass_models}.}
 \label{fig:rcz-mass_models}
\end{figure}

In this section, we study the evolution of lithium abundance predicted by the models, as a function of mass in each cluster, and as a function of age, and we compare the evolution between the three clusters. In Fig. \ref{fig:li-mass_models}, we present the lithium abundance as a function of mass for our model isochrones with ages that are estimated previously for each cluster. Continuous lines are the isochrones from models with extra mixing, whereas dotted lines are the isochrones from standard models. We can identify several interesting facts in this figure. Models with mass lower than $\sim1.10$ \msun, including rotation-induced mixing, seem to destroy their lithium early in the MS, but depletion continues throughout the main sequence, with a lower rate after Hyades age though. For 1.00 \msun, the lithium abundance difference between models of 1.59 Gyr for NGC 752 and models of 3.87 Gyr for M67 is about 0.5 dex. In this mass range, which is characterized by the presence of an extensive outer convective zone, lithium destruction is strongly dependent on stellar mass. 
We notice that the lithium depletion predicted at the Hyades age is greater than at the age of NGC 752 owing to the difference of metallicity between the models of the Hyades and the models of NGC 752. Indeed, the Hyades members are much more metallic than the NGC 752 members, and thus the models of the former have a deeper convective zone than  members of the latter (see Fig. \ref{fig:rcz-mass_models}), leading to a larger and quicker lithium depletion.
The mixing that occurs below that convective zone is mainly driven by rotation during the PMS and the beginning of the MS. However, surface lithium destruction during the PMS is significant only for the latest-type stars, and stars with masses around the solar mass present a slight lithium depletion at the ZAMS \citet{randich97}. In our models, PMS is not taken into account, but we calibrated our solar model to obtain the lithium abundance at the solar age, adopting the meteoritic lithium abundance of \citet{asplund09} as initial abundance. This implies that we probably used a stronger mixing than would have been necessary with a lower initial abundance owing to the slight depletion during the PMS. For lower masses, which should present a lower lithium abundance at the ZAMS, caused by the lithium destruction during the PMS, we expect lower lithium abundance than calculated by our models at the ages of the three clusters, and thus a 
steeper isochrone between 0.80 and 1.10 \msun.

For models with extra mixing with masses between 1.10 and 1.30 \msun, we observe the presence of a lithium plateau where the element is barely destroyed during evolution in the MS and independently of stellar mass. In these stars, the outer convective zone is very shallow and the turbulence is weak. The rotation-induced mixing below the convective zone does not sink deep enough to bring the lithium to the destruction layers. For these masses, lithium destruction during the PMS is negligible and it is consistent to use the same initial lithium abundance as for the solar model. Furthermore, a decrease in the mixing efficiency should not have any significant effect. The lithium abundance on this plateau seems to  slightly decrease with time, at a rate of about 0.2 dex between Hyades age and M67 age. 

When the mass increases, lithium abundance decreases with mass and age. This is due to the gravitational diffusion process below the convective zone. Lithium particles continuously sink below the convective zone owing to gravity and the amount of lithium in the convective zone decreases with time. For larger masses, the process is more efficient.

For the highest masses, we can see the dramatic decrease of lithium abundance with age due to the entry of the models in the red giant branch. When the star leaves the main sequence, the outer convective zone deepens during the first dredge-up and the lithium is diluted in lithium-free layers. In M67, this occurs in stars with masses around 1.30 \msun, although in NGC 752 this process can be observed for masses larger than 1.70 \msun.

Standard models have a totaly different behavior. For masses lower than about 1.25 \msun, depletion of lithium abundance is very weak and  is due to microscopic diffusion below the convective zone. When the mass increases, the outer convective zone shrinks and the diffusion is more efficient with time. For larger masses, the convective zone is very thin (see Fig. \ref{fig:rcz-mass_models}) and microscopic diffusion takes place on a timescale much smaller than the stellar life time on the main sequence. Surface lithium abundance is depleted very quickly. From comparison with models with extra mixing, we can deduce that the rotation-induced mixing prevents the lithium depeletion from microscopic diffusion in this range of masses, until the first dredge-up. These statements are long-standing results \citep[see][]{schatzman69}.

Figure \ref{fig:rcz-mass_models} shows the relative radius at the base of the convective zone along our three isochrones for the three open clusters and for the two sets of models as a function of stellar mass. The depth of the convection zone for the two sets of models is very similar, in contradiction with the completely different behaviour of the lithium depletion. However, in the case of the models with meridional circulation, the extra mixing lies below the base of the convective zone, increasing greatly the depth of the chemical transport. For low-mass stars, the outer convective zone becomes shallower when the mass increases. For mass lower than 1.10 \msun, this explains the mass-dependence of the lithium depletion. The convective zone of stars in the lithium plateau, i.e. between 1.10 and 1.30 \msun, is not deep enough for mixing mechanisms to reach the layers in which lithium could be destroyed. For larger masses, in the cases of the Hyades (M $>$ 1.60 \msun) and NGC 752 (1.40 \msun $<$ M $<$ 1.70 \msun), we can see that the slight slope in lithium depletion as a function of mass is not due to the depth of the convective zone, which remains nearly constant in this range of masses. As we explained, this lithium destruction is due to microscopic diffusion, which increases with mass and age. For larger masses, the dramatic decrease of lithium abundances corresponds to the strong deepening of the convective zone during the first dredge-up, at the beginning of the RGB. In the case of standard models, as explained above, the base of the convective zone is not deep enough to reach the lithium destruction layers, and the lithium depletion is controlled by the microscopic diffusion process, which is efficient for masses larger than about 1.25 \msun.

\section{Comparison of the predicted lithium abundance and observations}
\label{sec:comparisons}

\begin{figure}[t!]
 \begin{center}
 \vspace{-0.1in}
  \includegraphics[width=9 cm,height=9cm]{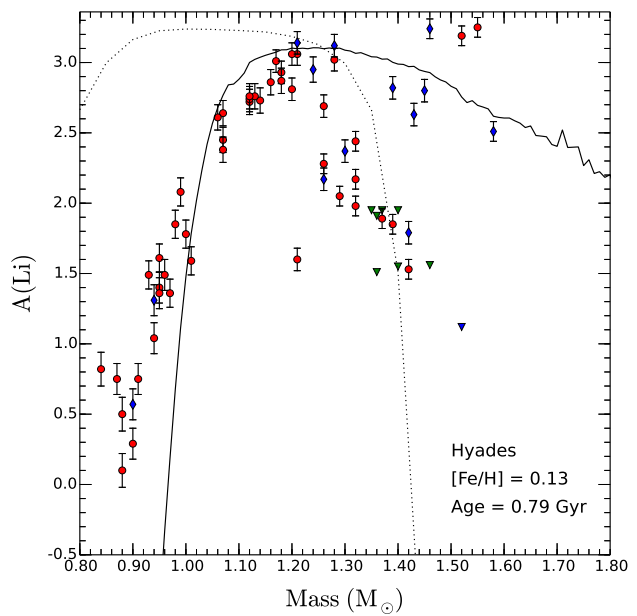}
 \end{center}
 \vspace{-0.2in}
 \caption{Lithium abundances of our sample as a function of the inferred stellar mass for the Hyades. Red filled circles represent stars with the lithium abundance determined. Green triangles represent the lithium abundance upper limits. Blue diamond represent the known binaries. Continuous line corresponds to the lithium abundance predicted by models with rotation-induced mixing at the age of the Hyades, whereas the dotted line corresponds to the standard models.}
 \label{fig:li-mass_hyades}
\end{figure}

For the \textbf{Hyades}, we estimated the mass for each star of our sample by picking the closest isochrone point to its data point on the color-magnitude diagram, and interpolating the value of the mass from the values for which the models were actually run. We used the isochrone calculated from models with extra mixing. Results of this interpolation are given in Table \ref{tab:interphyades}. In Figure \ref{fig:li-mass_hyades}, we present the lithium abundances of our sample as a function of the inferred stellar mass. We also plot the corresponding isochrones from both set of models. For stars with masses lower than 1.10 \msun, lithium destruction occurs as soon as the evolution starts. Stars with these masses present low lithium abundance at the age of the Hyades. Our models with extra mixing reproduce the depletion as observed in Fig. \ref{fig:li-mass_hyades}. The models present a more severe depletion for this mass range. Standard models do not reproduce to any extent the lithium abundances of these stars.

For this cluster, the lithium plateau at A(Li) $\sim 3.0$ for stars with masses larger than 1.10 \msun\ is also well reproduced. However, the Li-dip for stars with masses 
between 1.30 and 1.45 \msun\ is not reproduced by our models, thus suggesting that the mixing mechanism responsible for the Li-dip is not related with meridional circulation. However, it is interesting to note that the strong decrease of the lithium abundance in standard models corresponds to the Li-dip. For stars with masses larger than 1.60 \msun, there is a lack of lithium abundances estimation and we cannot give any formal comparison for the Hyades age. Known binaries have the same behavior as single stars. \\

\begin{figure}[t!]
 \begin{center}
 \vspace{-0.1in}
  \includegraphics[width=9 cm,height=9cm]{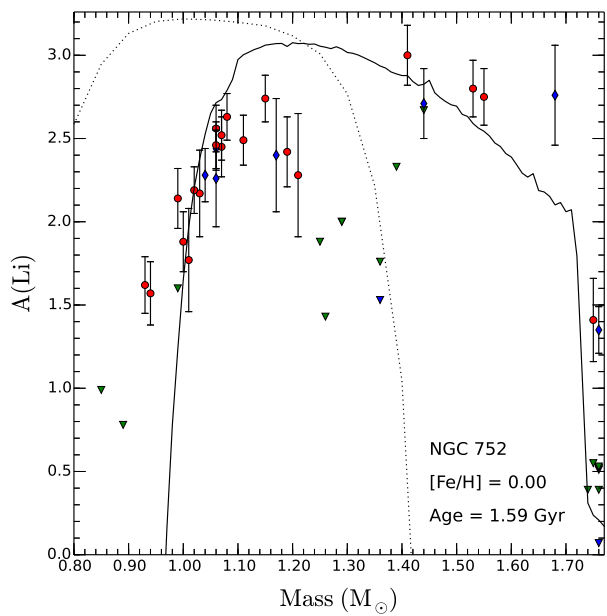}
 \end{center}
 \vspace{-0.2in}
 \caption{Same as Fig. \ref{fig:li-mass_hyades} but for the open cluster NGC 752.}
 \label{fig:li-mass_ngc752}
\end{figure}

For \textbf{NGC 752}, results of the interpolation are given in Table \ref{tab:interpngc752}. In Figure \ref{fig:li-mass_ngc752}, we plot the lithium abundances of our sample as a function of the inferred stellar mass for this cluster. We also present the corresponding isochrones from our two sets of models, with the ages deduced from the color-magnitude diagram. As in the Hyades, we observe a good agreement of the models with rotation-induced mixing for the stars of NGC 752 with masses that are lower than 1.10 \msun. However, for stars in the plateau, observed lithium abundances are lower by $\sim$0.3 dex in relation to to the models' predictions. The Li-dip is still not reproduced. The models with masses larger than 1.60 \msun \ seem to deplete too much lithium. Again, standard models do not reproduce the lithium distribution of the NGC 752 cluster, and the decrease of lithium abundance for more massive stars corresponds to the Li-dip. The lithium abundances of binaries have no 
pecularity compared to single stars. This implies that the binarity has no influence on lithium depletion of these stars and thus, and that there is no tidal effect in the observed systems. It suggests that there are no close binaries, and that an analysis of the spectra of the two components of each system should be done in
the future to know if they can be separated or not.\\

\begin{figure}[t!]
 \begin{center}
 \vspace{-0.1in}
  \includegraphics[angle=0,height=9cm,width=9cm]{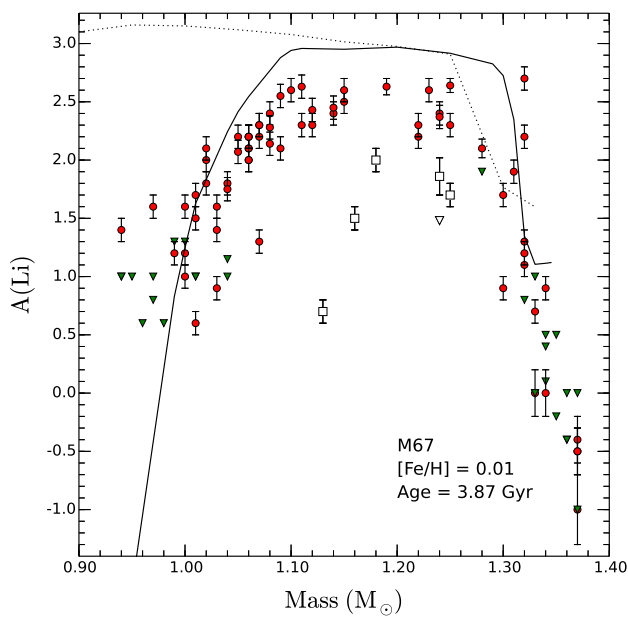}
 \end{center}
 \vspace{-0.2in}
 \caption{Same as Figs. \ref{fig:li-mass_hyades} and \ref{fig:li-mass_ngc752}, but for the open cluster M67. The open squares represents the deviant stars as discussed in the Sect. 5.3 of \citet{pace12}.}
 \label{fig:li-mass_m67}
\end{figure}

Figure \ref{fig:li-mass_m67} is the \textbf{M67} counterpart of Figs. \ref{fig:li-mass_hyades} and \ref{fig:li-mass_ngc752}. It is published by \citet{pace12} and shown here for better clarity and discussion. The results of the interpolation on the isochrone are also available in the online version of \citet{pace12}. In this paper, the authors observed a good agreement between models and observations for stars with masses lower than 1.10 \msun. The difference between the observed and the predicted plateaus is about 0.45 dex, showing an increase of this difference with age. The Li-dip, which lies around 1.34 \msun\ \citep{balachandran95}, is not entirely reproduced in depth, the decrease of lithium abundance in the models is only due to the deepening of the outer convective zone, when the star enters in the RGB. Extra mixing mechanisms responsible for the Li-dip are not yet identified.

\section{Discussion}
\label{sec:discussion}

Based on previous results with M67 \citep{pace12} and the new analysis for the Hyades and NGC 752, the calibration of evolution models on the atmospheric parameters of the Sun enables us to reproduce the qualitative distribution of lithium abundance as a function of the stellar mass for different ages. Lithium abundance is a crescent function of mass for $M < 1.10$ \msun, where the depth of the outer convective zone plays a crucial role in the efficiency of the meridional circulation to bring lithium from the base of the convective zone to the destruction layers. As expected, solar-calibrated mixing is suitable for solar-type stars at all stages of the evolution. For lower masses, our mixing seems to deplete too much lithium, too soon, and this discrepancy would be even larger if lithium depletion during the PMS had been taken into account.

For higher masses, we observe a plateau between 1.10 and 1.30 \msun in the three clusters\ where lithium remains constant as a function of stellar mass. This plateau is reproduced by our models. For these masses the outer convective zone is too shallow to allow the meridional circulation to bring lithium particles to the destruction layers. The lithium abundance in this plateau decreases with time, due to microscopic diffusion below the convective zone. Indeed, in Fig. \ref{fig:li-mass_models}, we can see that standard models present nearly the same lithium abundances as models with rotation-induced mixing in this range of masses. However, our models fail to reproduce quantitatively this decrease and the difference between the observed and the predicted plateaus increases with age. Since the turbulence, which is due to meridional circulation, prevents lithium to diffuse below the convective zone, it can indicate that our parametrization leads to a too strong efficiency of the meridional circulation for theses masses.

In the three clusters, we observe a Li-dip, which lies around 1.34 \msun. Our models with meridional circulation calibrated on the Sun do not reproduce this Li-dip. These results can be interpreted in different ways. As already stated, there is no a priori reason why the calibration of the parameters of the extra mixing for the solar model should also hold  for different masses, and at different ages. It has been shown and discussed by \citet{tv03b} and \citet{pace12} that it is possible to reproduce the Li-dip by calibrating the meridional circulation for each mass. We chose not to use this artificial parametric way and we focused on the study of the discrepancy in the solar mixing as a function of mass and its evolution. Furthermore, our results indicate that lithium depletion in these stars likely involves another mechanism besides meridional circulation. This other mechanism can be linked with angular momentum transport, like internal gravity waves \citep{talon&charbonnel05}. Internal gravity waves (IGW) 
are produced and excited by stochastic movement at the base of the convective zone and they transport mainly negative angular momentum through the radiative zone. They are mostly efficient in low-mass stars. If we introduce IGW in a solar model, part of the angular momentum should be transported by these waves, reducing the efficiency of the meridional circulation. Thus, we would have to increase the turbulent diffusion coefficients to reproduce the solar lithium depletion. This increase could be enough to destroy lithium in greater masses, where IGW are much less efficient, and reproduce the decrease of the lithium abundance in the plateau with time and the Li-dip. These new physics need to be tested for different masses. Moreover, it cannot be a coincidence that, for the three clusters, the Li-dip corresponds to the masses for which microscopic diffusion becomes very efficient. A possible explanation could be that, for some reason, the rotation-induced turbulence, which prevents microscopic diffusion being efficient in this range of masses, becomes frozen.
  
\begin{figure}[t!]
 \centering
 \vspace{-0.1in}
  \includegraphics[width=9cm,height=9cm]{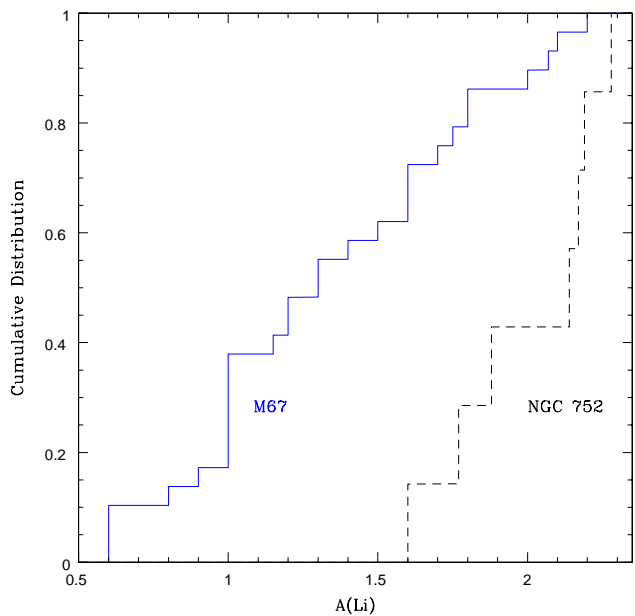}
 \vspace{-0.2in}
 \caption{Cumulative distribution for lithium abundance of NGC 752 (dotted line) stars and M67 (continuous blue line) stars with a stellar mass between 0.95 and 1.05 \msun.}
 \label{fig:hist_Li-mass}
\end{figure}

We are  also able to observe that for lower masses, the dispersion of lithium abundance seems to increase with age. The same trend is found for clusters studied by \citet{randich08} and \citet{randich09} for stars with a similar range of effective temperature. Figure \ref{fig:hist_Li-mass} shows the cumulative distribution of lithium abundance for a shallow range of mass around the solar value (between 0.95 and 1.05 \msun) in both open clusters with solar metallicity, NGC 752 and M67. Since the Hyades is overmetallic, it would produce an additional bias in  comparison. Indeed, a higher metallicity implies a deeper convective zone, and thus a larger efficiency of the transport mechanisms of lithium particles below the convective zone. In this figure, M67 presents a dispersion in A(Li) from 0.6 to 2.2 dex for masses between 0.95 and 1.05 \msun. NGC 752 presents a lower dispersion from 1.6 to 2.3 dex in the same range of masses. The dispersion seems to increase with age between 1.59 Gyr and 3.87 Gyr. However, the low 
number of stars with masses between 0.95 and 1.05 \msun \ for NGC 752 is not enough for a definitive conclusion about the evolution of the lithium abundance dispersion with age for solar-type stars. In \citet[][ in prep.]{duarte15}, the authors show that the same dispersion exists for solar twin field stars. The authors use a sample of 88 solar twin stars from \citet{ramirez14} and do a comparative study of lithium abundance, rotation period, and magnetic activity as a function of stellar mass. Lithium abundance in solar twin field stars span a wide range of values, which cannot be explained by the small differences in mass. This range of dispersion for stars with estimated ages of between 2.0 and 5.0 Gyr is   equivalent to the one we can see in the MS region in M67 ($M \leqslant 1.10$ \msun). 

It will be important to  understand this point in a near future, because it could be closely connected with the internal differential rotation evolution. Some studies put forward the idea that the dispersion in lithium abundances in solar-type stars is due to the different rotation history \citep[see][]{charbonneltalon05}. The way that the magnetic field of the protostar is coupled with the accretion disk determines the magnetic brake that the star undergoes during the PMS, and the initial rotation rate of the star at the ZAMS \citep[see][]{matt&pudritz05}. Furthermore, stellar winds have a strong influence on rotation in the MS, removing angular momentum from the surface convective zone of the star. This increases the differential rotation between the convective zone and the radiative interior, and thus the efficency of the tachocline, which plays a major role on lithium destruction. As a result, the resulting decrease in the rotation rate causes significant changes in the wind strength \citep{johnstone15}. 

This result indicates that lithium abundance is not such a good age indicator for solar-type stars, when used alone. Rotation period through gyrochronology \citep{barnes10} and asteroseismology must be taken into account to infer stellar age.

\section{Conclusions}
\label{sec:conclusions}

In this work, we have focused our study on the evolution of the atmospheric lithium abundance for open cluster, low-mass star members as a function of mass and age. We used three open clusters of different ages: the Hyades, with an estimated age of 0.79 Gyr, NGC 752, with an age of 1.59 Gyr, and M67, with 3.87 Gyr. For each cluster, we show how strong  the dependence of lithium abundance is with mass. Models that include meridional circulation reproduce most of the shape of lithium depletion. However, they fail to reproduce the Li-dip. Comparisons between clusters show that microscopic diffusion plays a role in F-type stars, where meridional circulation is inefficient in bringing lithium from the bottom of the shallow convective zone to the destruction layers. This study shows that there are two main reasons why using lithium abundance is tricky as an indicator of age. First, the lithium abundance shows a strong dependence on mass of a same age. The second reason is its dependency with transport mechanisms, which can 
be a function of rotation between others parameters. Comparisons of the rotation period distributions for solar-type stars in open clusters with different ages show that  core and  envelope rotations have to be coupled with a characteristic timescale of angular momentum transfer between them from a few Myr for the fastest rotators to nearly a hundred Myr for the slowest rotators \citep{denissenkov10,irwin07}. This angular momentum transfer seems to be mostly due to magnetic braking, but the interaction between meridional circulation and the differential rotation during this braking causes large turbulence which leads to a diffusion process of lithium from the base of the convective zone to the destruction layers. However, our models do not calculate this magnetic braking and the destruction of lithium is only driven by turbulence owing to classical meridional circulation. In the future, we intend to include the angular momentum redistribution in the stellar interior by magnetic braking in our models. 
Furthermore, in field stars one has to consider the intrinsic difficulty of getting accurate masses. This generates a greater dispersion and complicates the interpretation of the lithium abundance evolution as a function of age. However, for stars with very accurate mass determination, it is possible to use lithium abundance to diffentiate a young star from an old one. At the moment, we need observations of solar analog stars (with an effective temperature close to the solar one) of clusters with intermediate masses and solar metallicity to clarify the evolution of the dispersion of the lithium abundance in stars with solar mass.


\begin{acknowledgements}
This research made use of the SIMBAD database, operated at CDS, Strasbourg, France, and of WEBDA, an open cluster database developed and maintained by Jean-Claude Mermilliod. G.P. is supported by grant SFRH/BPD/39254/2007 and by the project PTDC/CTE-AST/098528/2008, funded by Funda\c{c}\~ao para a Ci\^encia e a Tecnologia (FCT), Portugal. Research activities of the  Stellar Board at the Federal University of Rio Grande do Norte are supported by continuous grants from CNPq and FAPERN Brazilian Agencies. J.D.N. and M.C. would like to acknowledge support from CNPq ({\em Bolsa de Produtividade}).
\end{acknowledgements}

\bibliographystyle{aa}

{}

\clearpage
\onecolumn

\setcounter{LTchunksize}{50}
\newcolumntype{C}{>{\centering\arraybackslash}m{8mm}}
\newcolumntype{D}{>{\centering\arraybackslash}m{10mm}}

\begin{longtable}{ C C C C D C C C D }
\caption{Results of the interpolation with the isochrone in the CMD for the Hyades star sample. First column: star ID, second column: effective temperature, third column: effective temperature of the closest point in the isochrone, fourth column: $\log g$, fifth column: $\log g$ of the closest point in the isochrone, sixth column: mass of the closest point in the isochrone, seventh column: lithium abundance, eighth column: lithium abundance error corresponding to an uncertainty
of 100 K in \teff, ninth column: star belonging (y) or not (n) to a binary system. \teff \ (col. 2), $\log g$ (col. 4), and the non-LTE lithium abundance $A({\rm Li})$ (col. 7) are from \citet{takeda13}.} \label{tab:interphyades}

\\

\hline 
\centering ID    & \teff & $T_{\rm iso}$ & $\log g$ & $(\log g)_{\rm iso}$ & Mass  & A(Li) & dA(Li) & Binarity\\  
\centering       & (K)   & (K)           &          &                      & \msun & (dex) & (dex) &  
 
\endfirsthead

\multicolumn{9}{c}%
{{\bfseries \tablename\ \thetable{} -- continued from previous page}} \\
\hline 
ID    & \teff & $T_{\rm iso}$ & $\log g$ & $(\log g)_{\rm iso}$ & Mass  & A(Li) & dA(Li) & Binarity\\  
      & (K)   & (K)           &          &                      & \msun & (dex) & (dex) & \\
\hline

\endhead

\hline 
\multicolumn{9}{l}{{Continued on next page}} \\ \hline
\endfoot

\\ 
\endlastfoot

\hline 

24357 &   6934 & 7000        &      4.30 &    4.30              &  1.52 &  3.19   & 0.07 & n\\
26462 &   6971 & 7014        &      4.30 &    4.19              &  1.55 &  3.25   & 0.07 & n\\
26015 &   6795 & 7079        &      4.32 &    4.18              &  1.58 &  2.51   & 0.07 & y\\
26911 &   6783 & 6823        &      4.32 &    4.24              &  1.46 &  3.24   & 0.07 & y\\
27561 &   6728 & 6823        &      4.33 &    4.24              &  1.46 &  <1.56  &      & n\\
18404 &   6714 & 6668        &      4.33 &    4.26              &  1.40 &  <1.55  &      & n\\
25102 &   6705 & 6606        &      4.33 &    4.28              &  1.37 &  <1.95  &      & n\\
28736 &   6693 & 6714        &      4.33 &    4.25              &  1.42 &  1.53   & 0.07 & n\\
26345 &   6660 & 6591        &      4.34 &    4.29              &  1.36 &  <1.51  &      & n\\
28568 &   6656 & 6591        &      4.34 &    4.29              &  1.36 &  <1.91  &      & n\\
28911 &   6651 & 6606        &      4.34 &    4.28              &  1.37 &  1.89   & 0.07 & n\\
27534 &   6598 & 6668        &      4.34 &    4.26              &  1.40 &  <1.95  &      & n\\
29225 &   6593 & 6561        &      4.35 &    4.29              &  1.35 &  <1.95  &      & n\\
27848 &   6558 & 6652        &      4.35 &    4.27              &  1.39 &  1.85   & 0.07 & n\\
31845 &   6558 & 6501        &      4.35 &    4.31              &  1.32 &  2.17   & 0.07 & n\\
28406 &   6554 & 6501        &      4.35 &    4.31              &  1.32 &  2.44   & 0.07 & n\\
27483 &   6533 & 6934        &      4.36 &    4.20              &  1.52 &  <1.12  &      & y\\
27731 &   6507 & 6426        &      4.36 &    4.33              &  1.29 &  2.05   & 0.07 & n\\
28483 &   6474 & 6501        &      4.37 &    4.31              &  1.32 &  1.98   & 0.07 & n\\
28608 &   6465 & 6353        &      4.37 &    4.35              &  1.26 &  2.28   & 0.07 & n\\
30869 &   6339 & 6714        &      4.39 &    4.25              &  1.42 &  1.79   & 0.08 & y\\
27383 &   6310 & 6441        &      4.40 &    4.33              &  1.30 &  2.37   & 0.08 & y\\
27991 &   6310 & 6792        &      4.40 &    4.24              &  1.45 &  2.80   & 0.08 & y\\
26784 &   6291 & 6397        &      4.40 &    4.34              &  1.28 &  3.02   & 0.08 & n\\
27808 &   6275 & 6223        &      4.41 &    4.39              &  1.21 &  3.06   & 0.08 & n\\
28394 &   6242 & 6353        &      4.41 &    4.35              &  1.26 &  2.17   & 0.08 & y\\
30809 &   6239 & 6223        &      4.41 &    4.39              &  1.21 &  1.60   & 0.08 & n\\
28363 &   6202 & 6745        &      4.42 &    4.25              &  1.43 &  2.63   & 0.08 & y\\
30738 &   6202 & 6397        &      4.42 &    4.34              &  1.28 &  3.12   & 0.08 & y\\
28205 &   6199 & 6194        &      4.42 &    4.40              &  1.20 &  3.06   & 0.08 & n\\
28635 &   6186 & 6123        &      4.42 &    4.42              &  1.17 &  3.01   & 0.08 & n\\
30810 &   6174 & 6652        &      4.43 &    4.27              &  1.39 &  2.82   & 0.08 & y\\
35768 &   6122 & 6353        &      4.44 &    4.35              &  1.26 &  2.69   & 0.08 & n\\
28033 &   6118 & 6223        &      4.44 &    4.39              &  1.21 &  3.14   & 0.08 & y\\
27406 &   6107 & 6137        &      4.44 &    4.41              &  1.18 &  2.93   & 0.08 & n\\
28237 &   6107 & 6194        &      4.44 &    4.40              &  1.20 &  2.81   & 0.08 & n\\
20430 &   6079 & 6309        &      4.45 &    4.37              &  1.24 &  2.95   & 0.09 & y\\
14127 &   6079 & 5984        &      4.45 &    4.46              &  1.12 &  2.72   & 0.09 & n\\
29419 &   6045 & 6095        &      4.45 &    4.43              &  1.16 &  2.86   & 0.09 & n\\
30589 &   6037 & 6137        &      4.45 &    4.41              &  1.18 &  2.87   & 0.09 & n\\
25825 &   5983 & 5984        &      4.46 &    4.46              &  1.12 &  2.74   & 0.09 & n\\
27859 &   5961 & 6039        &      4.47 &    4.44              &  1.14 &  2.73   & 0.09 & n\\
28344 &   5924 & 6011        &      4.47 &    4.45              &  1.13 &  2.76   & 0.09 & n\\
20439 &   5894 & 5984        &      4.48 &    4.46              &  1.12 &  2.76   & 0.09 & n\\
28992 &   5844 & 5834        &      4.49 &    4.49              &  1.07 &  2.64   & 0.09 & n\\
26767 &   5812 & 5794        &      4.50 &    4.50              &  1.06 &  2.61   & 0.09 & n\\
26736 &   5757 & 5834        &      4.51 &    4.49              &  1.07 &  2.45   & 0.09 & n\\
28099 &   5735 & 5834        &      4.51 &    4.49              &  1.07 &  2.38   & 0.09 & n\\
26756 &   5640 & 5571        &      4.53 &    4.55              &  0.99 &  2.08   & 0.10 & n\\
27282 &   5553 & 5610        &      4.54 &    4.54              &  1.00 &  1.78   & 0.10 & n\\
240648&   5527 & 5533        &      4.54 &    4.55              &  0.98 &  1.85   & 0.10 & n\\
19902 &   5522 & 5636        &      4.55 &    4.53              &  1.01 &  1.59   & 0.10 & n\\
28593 &   5516 & 5508        &      4.55 &    4.56              &  0.97 &  1.36   & 0.10 & n\\
31609 &   5508 & 5432        &      4.55 &    4.57              &  0.95 &  1.61   & 0.10 & n\\
27250 &   5485 & 5357        &      4.55 &    4.58              &  0.93 &  1.49   & 0.10 & n\\
27732 &   5449 & 5395        &      4.55 &    4.58              &  0.94 &  1.31   & 0.11 & y\\
32347 &   5429 & 5470        &      4.56 &    4.57              &  0.96 &  1.49   & 0.11 & n\\
242780&   5429 & 5432        &      4.56 &    4.57              &  0.95 &  1.40   & 0.11 & n\\
283704&   5426 & 5432        &      4.56 &    4.57              &  0.95 &  1.36   & 0.11 & n\\
284574&   5303 & 5395        &      4.57 &    4.58              &  0.94 &  1.04   & 0.11 & n\\
284253&   5297 & 5272        &      4.57 &    4.59              &  0.91 &  0.75   & 0.11 & n\\
285773&   5254 & 5116        &      4.58 &    4.62              &  0.87 &  0.75   & 0.11 & n\\
30505 &   5249 & 5236        &      4.58 &    4.60              &  0.90 &  0.57   & 0.11 & y\\
28258 &   5235 & 5236        &      4.58 &    4.60              &  0.90 &  0.29   & 0.11 & n\\
27771 &   5196 & 5164        &      4.58 &    4.61              &  0.88 &  0.50   & 0.12 & n\\
28462 &   5172 & 4988        &      4.58 &    4.63              &  0.84 &  0.82   & 0.12 & n\\
285367&   5114 & 5164        &      4.59 &    4.61              &  0.88 &  0.10   & 0.12 & n\\
\end{longtable}


\newpage

\begin{longtable}{ D C D C C D C C C D }
\caption{Results of the interpolation with the isochrone in the CMD for the NGC 752 star sample.First column: star ID, second column: effective temperature of the closest point in the isochrone, third column: $\log g$ of the closest point in the isochrone, fourth column: photometric temperature, fifth column: mass of the closest point in the isochrone, sixth column: equivalent width of the Li line at 6708 \AA, seventh column: equivalent width error, eighth column: lithium abundance, ninth column: lithium abundance error, tenth column: star belonging (y) or not (n) to a binary system. EW measurements derived from published lithium abundances are flagged with an asterisk.} \label{tab:interpngc752}

\\

\hline 
ID      & $T_{\rm iso}$ & $(\log g)_{\rm iso}$ & $T_{\rm phot}$ & Mass  & EW$_{Li}$ & dEW$_{Li}$ & A(Li) & dA(Li) & Binarity\\  
        & (K)           &                      & (K)            & \msun & (m$\AA$)         &(m$\AA$)           & (dex) & (dex)  & \\
 
\endfirsthead

\multicolumn{10}{c}%
{{\bfseries \tablename\ \thetable{} -- continued from previous page}} \\
\hline 
ID      & $T_{\rm iso}$ & $(\log g)_{\rm iso}$ & $T_{\rm phot}$ & Mass  & EW$_{Li}$ & dEW$_{Li6708\AA}$ & A(Li) & dA(Li) & Binarity\\  
        & (K)           &                      & (K)            & \msun & (m$\AA$)  &(m$\AA$)           & (dex) & (dex)  & \\
\hline

\endhead

\hline 
\multicolumn{10}{c}{{Continued on next page}} \\ \hline
\endfoot

\\ 
\endlastfoot

\hline 

H10     &    6854       &   3.82               & 6463           & 1.68  &     35  &    10  & 2.69  & 0.30 & y\\    
H88     &    6501       &   4.29               & 6195           & 1.29  &    <10  &        &<2.00  &      & n\\    
H94     &    5308       &   4.59               & 5333           & 0.89  &     <5  &        &<0.78  &      & n\\    
H123    &    6668       &   4.14               & 6523           & 1.44  &    <34  &        &<2.67  &      & n\\    
H139    &    6501       &   4.29               & 6294           & 1.29  &    <10  &        &<2.00  &      & n\\    
H185    &    6266       &   4.36               & 6030           & 1.19  &     42  &    10  & 2.31  & 0.21 & n\\    
H189    &    6668       &   4.17               & 6416           & 1.41  &     65  &    10  & 3.00  & 0.18 & n\\    
H222    &    6652       &   4.03               & 6555           & 1.55  &     45  &    10  & 2.68  & 0.17 & n\\    
H227    &    5116       &   4.62               & 4917           & 0.85  &    <12  &        &<0.99  &      & n\\    
H235    &    6606       &   4.21               & 6203           & 1.36  &     <3  &        &<1.53  &      & y\\    
H237    &    6223       &   4.38               & 5797           & 1.17  &     43  &    10  & 2.29  & 0.34 & y\\    
H254    &    6668       &   4.05               & 6617           & 1.53  &     49  &    10  & 2.74  & 0.17 & n\\    
H259    &    6637       &   4.19               & 6432           & 1.39  &    <17  &        &<2.33  &      & n\\    
H265    &    5688       &   4.53               & 5614           & 0.99  &    <15  &        &<1.60  &      & n\\    
H266    &    6668       &   4.15               & 6412           & 1.44  &     42  &    10  & 2.64  & 0.21 & y\\    
H293    &    6412       &   4.31               & 6188           & 1.25  &    <15  &        &<1.32  &      & n\\    
H302    &    6606       &   4.22               & 6440           & 1.36  &     <5  &        &<1.76  &      & n\\    
H304    &    6441       &   4.30               & 6199           & 1.26  &     <3  &        &<1.43  &      & n\\    
PLA475  &    5929       &   4.47               & 5722           & 1.07  &     59  &    10  & 2.45  & 0.18 & n\\    
PLA520  &    5929       &   4.47               & 5893           & 1.07  &     67  &    12  & 2.52  & 0.15 & n\\    
PLA552  &    5888       &   4.48               & 5857           & 1.06  &     61  &     7  & 2.44  & 0.13 & y\\    
PLA648  &    6324       &   4.35               & 5844           & 1.21  &     24  &     8  & 2.28  & 0.37 & n\\    
PLA699  &    5847       &   4.50               & 5713           & 1.04  &     48  &    10  & 2.28  & 0.16 & y\\    
PLA701  &    5794       &   4.50               & 5526           & 1.03  &     42  &    12  & 2.17  & 0.26 & n\\    
PLA786  &    5727       &   4.52               & 5414           & 1.01  &     20  &     7  & 1.77  & 0.31 & n\\    
PLA859  &    5714       &   4.52               & 5584           & 1.00  &     26  &     7  & 1.88  & 0.18 & n\\    
PLA864  &    5902       &   4.47               & 5867           & 1.06  &     75  &     9  & 2.56  & 0.14 & n\\    
PLA889  &    5970       &   4.46               & 5956           & 1.08  &     78  &    11  & 2.63  & 0.14 & n\\    
PLA921  &    6053       &   4.44               & 5949           & 1.11  &     54  &     9  & 2.49  & 0.15 & n\\    
PLA964  &    5888       &   4.48               & 5854           & 1.06  &     64  &     9  & 2.46  & 0.14 & n\\    
PLA983  &    5767       &   4.51               & 5765           & 1.02  &     45  &     7  & 2.19  & 0.14 & n\\    
PLA993  &    5482       &   4.57               & 5475           & 0.94  &     20  &     6  & 1.57  & 0.19 & n\\    
PLA1012 &    6165       &   4.40               & 5996           & 1.15  &     74  &    10  & 2.74  & 0.14 & n\\    
PLA1107 &    5445       &   4.58               & 5614           & 0.93  &     24  &     5  & 1.62  & 0.17 & n\\    
PLA1284 &    5902       &   4.48               & 5564           & 1.06  &     42  &     9  & 2.26  & 0.29 & y\\    
PLA1365 &    5662       &   4.54               & 5470           & 0.99  &     48  &     8  & 2.14  & 0.18 & n\\    
H77     &    4764       &   2.93               & 4580           & 1.75  &     59* &    10  & 1.41  & 0.25 & n\\    
H311    &    4677       &   2.75               & 4718           & 1.76  &    <12* &        &<0.52  &      & n\\    
H137    &    4634       &   2.68               & 4773           & 1.76  &    <10* &        &<0.39  &      & n\\    
H208    &    4623       &   2.65               & 4682           & 1.76  &     70* &    10  & 1.35  & 0.14 & y\\    
H213    &    4666       &   2.73               & 4812           & 1.76  &    <12* &        &<0.51  &      & y\\    
H1      &    4764       &   2.93               & 4934           & 1.75  &    <10* &        &<0.55  &      & n\\    
H295    &    4731       &   2.87               & 4891           & 1.76  &    <10* &        &<0.52  &      & n\\    
H186    &    4931       &   3.30               & 4891           & 1.74  &     <5* &        &<0.39  &      & n\\    
H110    &    4645       &   2.70               & 5153           & 1.76  &     <5* &        &<0.07  &      & y\\    
H24     &    4645       &   2.69               & 4812           & 1.76  &    <13* &        &<0.53  &      & n\\    
\end{longtable}

\end{document}